\documentstyle[12pt]{article}
\newlength{\abstractwidth}
\renewcommand{\thanks}[1]{\footnote{#1}} % Use this for footnotes

\newcommand{\be}{\begin{equation}}
\newcommand{\bea}{\begin{eqnarray}}
\newcommand{\eea}{\end{eqnarray}}
\newcommand{\ee}{\end{equation}}

\newcommand{\<}{\langle}
\renewcommand{\>}{\rangle}
\def\ba{\begin{eqnarray}}
\def\ea{\end{eqnarray}}

\def\A{{\cal A}}

\def\O{{\cal O}}

\def\X{{\cal X}}
\def\Z{{\cal Z}}

\def\det{{\rm det}}

\def\half{ {1\over 2}}
\def\third{{1 \over 3}}
\def\p{\partial}
\def\pz{\partial _z}

\def\pw{\partial _w}

\def\tet{\vartheta}

\def\chiz{{\chi _{\bar z}{} ^+}}
\def\chiw{{\chi _{\bar w}{} ^+}}
\def\chiu{{\chi _{\bar u}{} ^+}}
\def\chiv{{\chi _{\bar v}{} ^+}}

\begin{document}
\baselineskip=16pt

\begin{flushright}
UCLA/01/TEP/26 \\
Columbia/Math/01
\end{flushright}

\bigskip

\begin{center}
{\Large \bf TWO-LOOP SUPERSTRINGS \ III }
 \\
\bigskip
{\large \bf Slice Independence and Absence of Ambiguities
\footnote{Research supported in
part by National Science Foundation grants PHY-98-19686 and DMS-98-00783,
and by the Institute for Pure and Applied Mathematics under NSF grant
DMS-9810282.}}

\bigskip\bigskip

{\large Eric D'Hoker$^a$ and D.H. Phong$^b$} \\ 

\bigskip

$^a$ \sl Department of Physics and \\
\sl Institute for Pure and Applied Mathematics (IPAM) \\
\sl University of California, Los Angeles, CA 90095 \\
$^b$ \sl Department of Mathematics \\ 
\sl Columbia University, New York, NY 10027

\end{center}

\bigskip\bigskip

\begin{abstract}

The chiral superstring measure constructed in the earlier papers of 
this series for general gravitino slices $\chiz$ is examined in 
detail for slices supported at two points $x_1$ and $x_2$, $\chiz = \zeta
^1 \delta (z,x_1)  + \zeta ^2 \delta (z,x_2)$, where $\zeta ^1$ and
$\zeta^2$ are the odd Grassmann valued supermoduli. In this case, the
invariance of the measure under infinitesimal changes of gravitino slices
established previously is strengthened to its most powerful form: the
measure is shown, point by point on moduli space, to be locally and
globally independent from $x_{\alpha}$, as well as from the superghost
insertion points $p_a$, $q_{\alpha}$  introduced earlier as computational
devices. In particular, the measure is completely unambiguous. The limit
$x_{\alpha}=q_{\alpha}$ is then well defined. It is of special interest,
since it elucidates some subtle issues in the construction of the
picture-changing operator $Y(z)$ central to the BRST formalism. The
formula for the chiral superstring measure in this limit is derived
explicitly.

\end{abstract}

\vfill\eject

\section{Introduction}
\setcounter{equation}{0}

This paper is the third of a series whose goal is to show that two-loop
amplitudes in superstring theory are fully slice-independent, do not
suffer from any ambiguity, and can actually be expressed explicitly in
terms of modular forms and sections of vector bundles over the moduli
space of Riemann surfaces. The main formulas obtained have been announced
in \cite{I} (hereafter referred to as I), with the full derivations to
appear in subsequent papers.  In \cite{II} (hereafter referred to as II),
the first step has  been carried out in detail, which is to derive from
first principles a formula for the gauge-fixed amplitude, and to establish
its invariance under  infinitesimal changes of gauge slices.   

\medskip

The main purpose of the present paper III is to consider the family of
worldsheet gravitino slices with support at two points $x_\alpha$,
$\alpha =1,2$,
\bea
\label{point}
\chiz = \sum _{\alpha =1,2}   \zeta ^\alpha \delta (z,x_\alpha) \, 
\eea
and to prove the full-fledged invariance, both under infinitesimal as well
as global changes, of  the gauge-fixed formula when restricted to this
family. Gravitino slices of the form (\ref{point}) are of great practical
interest, since the gauge-fixed formula can be expressed then
most simply in terms of meromorphic Green's functions and
holomorphic differentials evaluated at $x_\alpha$
and the auxiliary superghost insertions $q_\alpha$ and $p_a$,
$a=1,2,3$. Our principal result in III  
is that the formula thus obtained is in fact {\sl locally and
globally independent of all the points $x_\alpha$, $q_\alpha$ and $p_a$.}
The independence from the points
$q_\alpha$ and $p_a$ was expected since these points were introduced
merely as a computational device. The additional proof of their
independence can be viewed as a check on the consistency of the
entire approach and final formula. The independence from
the points $x_\alpha$ is the crucial new fact,
which really constitutes a proof
that the chiral superstring measure is unambiguous and 
globally slice independent.

\medskip

The independence from the points $x_{\alpha}$, $q_{\alpha}$, and $p_a$
leads to a simpler gauge-fixed formula, where we can set $x_{\alpha}\to
q_{\alpha}$. This simpler formula is also more natural, since the points
$p_a$ can be viewed as a slice for moduli, and the points $q_{\alpha}$ as
a slice for supermoduli. More important, in the limit $x_{\alpha}\to
q_{\alpha}$, the supercurrent insertions $S(x_\alpha)$ are made to
coincide with the superghost insertions $\delta (\beta (q_\alpha))$.
Formally, as shown in \cite{fms,vv87}, this product should yield the
picture-changing operator $ Y(z)=\delta(\beta(z))S(z)$ although the
difficulties associated with taking $\delta(\beta(z))$ and $S(z)$ at
coincident points had not been clarified before. A byproduct of our 
analysis is that indeed, the naive product of the  supercurrent and the
superghost insertions is then singular and ill-defined (c.f. Section 3.3
below). Including the subtle new contributions of our gauge-fixing
procedure, however, the limit is automatically well-defined.

\medskip

We obtain in this way what amounts to the correct(ed) prescription
replacing that of the picture changing operator. The final formula is the
main result of this paper, 
\bea
\label{finamppq}
{\cal A} [\delta]
& = & i \Z 
\biggl \{ 1  + {\cal X}_1 + {\cal X}_2 + {\cal X}_3 + {\cal X}_4 +  {\cal
X}_5 + {\cal X}_6 \biggr \}
\nonumber \\
\Z & = & {\< \prod _a b(p_a) \prod _\alpha \delta (\beta (q_\alpha))
\> \over \det \omega _I \omega _J (p_a) }
\eea 
and the $\X_i$ are given by
\bea 
\label{peeque}
\X_1 + \X _6 &=&
{\zeta ^1 \zeta ^2 \over 16 \pi ^2}
\biggl [
-10 S_\delta (q_1,q_2) \p _{q_1} \p _{q_2} \ln E(q_1,q_2)
 \\
&&
\qquad \
 - \p _{q_1} G_2 (q_1,q_2) \p \psi ^* _1 (q_2)
 + \p _{q_2} G_2 (q_2,q_1) \p \psi ^* _2 (q_1)
\nonumber \\
&&
\qquad \
+ 2   G_2 (q_1,q_2) \p \psi ^* _1 (q_2)  f_{3/2} ^{(1)} (q_2)
- 2   G_2 (q_2,q_1) \p \psi ^* _2 (q_1)  f_{3/2} ^{(2)} (q_1)
\biggr ]
\nonumber \\
\X _2 &=&
{\zeta ^1 \zeta ^2 \over 16 \pi ^2}
\omega _I(q_1) \omega _J(q_2) S_\delta (q_1,q_2) 
\biggl [  \p _I \p _J \ln {\tet [\delta ](0)^5 \over \tet
[\delta ](D_\beta )} +   \p _I \p _J \ln \tet (D_b ) \biggr ]
\nonumber \\
\X _3 &=&
{\zeta ^1 \zeta ^2 \over 8 \pi ^2}
S_\delta (q_1,q_2) \sum _a  \varpi _a  (q_1, q_2) \biggl [ B_2(p_a) +
B_{3/2}(p_a) \biggr ]
\nonumber \\
\X_4 &=& {\zeta ^1 \zeta ^2 \over 8 \pi ^2}
S_\delta (q_1,q_2) \sum _a \biggl [
\p _{p_a} \p _{q_1} \ln E(p_a,q_1) \varpi ^* _a(q_2)
+ \p _{p_a} \p _{q_2} \ln E(p_a,q_2) \varpi ^* _a(q_1) \biggr ]
\nonumber \\
\X _5 &=&
{\zeta ^1 \zeta ^2 \over 16 \pi ^2}
\sum _a \biggl [ 
S_\delta (p_a, q_1) \p _{p_a} S_\delta (p_a,q_2) 
- S_\delta (p_a, q_2) \p _{p_a} S_\delta (p_a,q_1) \biggr ] 
\varpi _a  (q_1,q_2) \, .
\nonumber
\eea
Here, the quantity $\p \psi ^* _1 (q_2)$ is a tensor, given by 
\be
\p \psi ^* _1 (q_2) =
{\tet [\delta ] (q_2 -q_1 + D_\beta ) \over \tet [\delta ] (D_\beta)
E(q_1,q_2)}{ \sigma (q_2) ^2 \over \sigma (q_1)^2}\, ,
\ee
and we have 
\bea
\label{effs}
f_n(w) & = &
\omega_I(w) \p_I \ln \tet [\delta](D_n)
+\p_w \ln \bigg ( \sigma(w)^{2n-1}\prod_{i=1}^{2n-1}E(w,z_i) \biggr )
 \\
f_{3/2} ^{(1)} (x) &=&
\omega _I (q_1) \p _I \ln \tet [\delta ] (x + q_2 - 2 \Delta)
+ \p _{q_1} \ln \biggl (  E(q_1,q_2) E(q_1,x) \sigma (q_1)^2 \biggr )
\nonumber \\
f_{3/2} ^{(2)} (x) &=&
\omega _I (q_2) \p _I \ln \tet [\delta ] (x + q_1 - 2 \Delta)
+ \p _{q_2} \ln \biggl (  E(q_2,q_1) E(q_2,x) \sigma (q_2)^2 \biggr )\, .
\nonumber
\eea
The quantities $B_2$ and $B_{3/2}$ are defined as follows,
\bea
B_2 (w) & = & -27 T_1(w) +
 \half f_2 (w)^2 -{3 \over 2} \pw f_2(w)-2 \sum _a \p_{p_a} \pw \ln
E(p_a,w) \varpi ^* _a (w)
\nonumber \\
B_{3/2} (p_a) & = & 12 T_1 (p_a) - \half f_{3/2}(p_a)^2 + \p f_{3/2}(p_a)
\eea
All other quantities in (\ref{finamppq}) and (\ref{peeque}) were defined
in II, and will be discussed again below. Further background material
can be found in \cite{superm,dp88,dp89}.

\vfill\eject

\section{Gravitino Slices Supported at Points}
\setcounter{equation}{0}

The starting point of this paper is the two-loop, even spin structure,
chiral superstring measure derived in paper II for an arbitrary gravitino
gauge slice 
\be
\chiz=\sum_{\alpha=1}^2\zeta^{\alpha} (\chi_{\alpha}) _{\bar z}{}^+
\ee 
It is given by the following expression,
\bea
\label{finamp}
{\cal A} [\delta]
& = &
i \ {\cal Z}
\biggl \{ 1  + {\cal X}_1 + {\cal X}_2 + {\cal X}_3 + {\cal X}_4 +  {\cal
X}_5 + {\cal X}_6 \biggr \}
\nonumber \\
{\cal Z}
& = &
{\< \prod _a b(p_a) \prod _\alpha \delta (\beta (q_\alpha)) \>
\over \det \bigl (\omega _I \omega _J (p_a) \bigr )
 \cdot \det \< \chi _\alpha | \psi ^* _\beta\>}
\eea
with the $\X_i$ defined as follows 
\bea
\label{Xes}
{\cal X}_1
&=&
 - {1 \over 8 \pi ^2} \int \! d^2z \chiz \int \! d^2 w \chiw \< S(z) S(w)\> 
\nonumber \\
{\cal X} _2 
& = & + {i \over 4\pi} (\hat \Omega _{IJ} - \Omega _{IJ} )
\biggl (  5\p _I \p _J \ln \tet [\delta ](0) - \p _I \p _J \ln \tet
[\delta ](D_\beta ) +   \p _I \p _J \ln \tet (D_b ) \biggr )
\nonumber \\
{\cal X} _3 
&=& + {1 \over 2 \pi} \int d^2 z \hat \mu (w)  \biggl ( B_2(w) +
B_{3/2}(w)
\biggr )
\nonumber \\
{\cal X} _4 
&=& + {1 \over 8\pi ^2} \int \! d^2w \ \p _{p_a} \pw \ln E(p_a,w) \chiw 
\int \! d^2u S_\delta (w,u) \chiu \varpi ^* _a(u)
\nonumber \\
{\cal X} _5 
&=& + {1 \over 16 \pi ^2} \int \! d^2u \int \! d^2v S_\delta (p_a,u) \chiu  
\p _{p_a} S_\delta (p_a,v) \chiv \varpi _a  (u,v) 
\nonumber \\
{\cal X} _6
&=& + {1 \over 16 \pi ^2} \int \! d^2z \chi _\alpha ^* (z) \int \!
d^2w G_{3/2} (z,w) \chiw \int \! d^2 v \chiv \Lambda _\alpha (w,v)  
\eea
Explicit formulas are available in Appendix A of \cite{II}
for all the ingredients of this formula, such as the Green's functions for
$b,c$ ghosts $G_2(z,w)$ and for $\beta, \gamma$ superghosts $G_{3/2}(z,w)$, for the
prime form $E(z,w)$, for the Szeg\" o kernel $S_\delta(z,w)$, and for the
holomorphic 3/2 differentials $\psi ^* _\alpha$, normalized at points
$q_\beta$ by $\psi ^* _\alpha (q_\beta) = \delta _{\alpha \beta}$. The
Beltrani differential $\hat \mu$ effects the deformation of complex
structures from the super period matrix to the supergeometry of the slice.
We shall not repeat those definitions here, but refer the reader to
\cite{II} instead. There is an explicit formula available for the
supercurrent correlator,
\bea
\< S(z) S(w) \> 
&=&
+ {5 \over 2} \p_z \p_w \ln E(z,w) S_\delta (z,w) 
\nonumber \\ &&
+ {3 \over 4} \p_w G_2 (z,w) G_{3/2} (w,z) + \half G_2 (z,w) \p_w G_{3/2}
(w,z)
\nonumber \\ &&
- {3 \over 4} \p_z G_2 (w,z) G_{3/2} (z,w) - \half G_2 (w,z) \p_z G_{3/2}
(z,w) \, .
\eea
Furthermore, $B_2$ and $B_{3/2}$ are holomorphic two forms, which are
given by
\bea
\label{beetwo}
B_2(w) = -27 T_1(w) +
 \half f_2 (w)^2 -{3 \over 2} \pw f_2(w)-2 \sum _a \p_{p_a} \pw \ln
E(p_a,w) \varpi ^* _a (w) \, .
\eea
\bea
\label{bthreehalfs}
B_{3/2}(w) &=& 12 T_1(w) 
-\half f_{3/2}(w)^2 + \pw f_{3/2}(w) 
 \\
&& +\int \! d^2z \chi ^* _\alpha (z) \biggl (
-{3 \over 2} \pw G_{3/2} (z,w) \psi ^* _\alpha (w)
  -\half  G_{3/2} (z,w) \pw \psi ^* _\alpha (w)
\nonumber \\
&& \qquad \qquad \qquad \qquad 
+ G_2 (w,z) \pz \psi ^* _\alpha (z) + {3 \over 2} \pz G_2 (w,z) \psi
^* _\alpha (z) \biggr )
\nonumber
\eea
Finally, $\chi ^* _\alpha$ are the linear combinations of $\chi _\alpha$
normalized so that $\< \chi ^* _\alpha |\psi ^* _\beta \> = \delta
_{\alpha \beta}$ and the expression $\Lambda _\alpha$ is given by
\be
\label{Lambda}
 \Lambda _\alpha (w,v) 
=
2 G_2(w,v) \p _v \psi _\alpha ^* (v) 
+ 3 \p _v G_2 (w,v)  \psi _\alpha ^* (v)
\ee

\subsection{The $\delta$-function gravitino expressions for $\X_1$,
$\X_2$, $\X_3$, $\X_4$, $\X_5$}

Considerable simplifications take place when the choice $\chi _\alpha
(w) = \delta (w,x_\alpha)$ is made, even for points $x_\alpha$ that are
arbitrary  and unrelated to $p_a$ or $q_\alpha$. The limits of 
$\X_i$, $i=1,\cdots ,5$ are manifestly regular, while that of $\X _6$ is
regular only after careful manipulations.

We may readily evaluate the overall factor $\Z$ and the terms $\X_i$,
$i=1,\cdots ,5$ that enter into (\ref{finamp}) and (\ref{Xes}) on the
gravitino slice $\chi _\alpha (w) = \delta (w, x _\alpha)$, and we find
\bea
\label{Zformula}
\Z = {\< \prod _a b(p_a) \prod _\alpha \delta (\beta (q_\alpha))
\> \over \det \omega _I \omega _J (p_a) \cdot \det \psi ^* _\beta
(x_\alpha) }
\eea
\bea
\label{regularX}
\X_1 &=&
{\zeta ^1 \zeta ^2 \over 16 \pi ^2}
\biggl [
-10 S_\delta (x_1,x_2) \p _{x_1} \p _{x_2} \ln E(x_1,x_2)
\nonumber \\
&&
\qquad \ \
-3 \p _{x_2} G_2 (x_1, x_2) G_{3/2}(x_2,x_1) - 2 G_2 (x_1,x_2) \p _{x_2}
G_{3/2}(x_2,x_1) 
\nonumber \\
&&
\qquad \ \
 +3 \p _{x_1} G_2 (x_2, x_1) G_{3/2}(x_1,x_2) + 2 G_2 (x_2,x_1) \p _{x_1}
G_{3/2}(x_1,x_2)
\biggr ]
\nonumber \\
\X _2 &=&
{\zeta ^1 \zeta ^2 \over 16 \pi ^2}
\omega _I(x_1) \omega _J(x_2) S_\delta (x_1,x_2) 
\biggl [  \p _I \p _J \ln {\tet [\delta ](0)^5 \over \tet
[\delta ](D_\beta )} +   \p _I \p _J \ln \tet (D_b ) \biggr ]
\nonumber \\
\X _3 &=&
{\zeta ^1 \zeta ^2 \over 8 \pi ^2}
S_\delta (x_1,x_2) \sum _a  \varpi _a  (x_1, x_2) \biggl [ B_2(p_a) +
B_{3/2}(p_a) \biggr ]
 \\
\X_4 &=& {\zeta ^1 \zeta ^2 \over 8 \pi ^2}
S_\delta (x_1,x_2) \sum _a \biggl [
\p _{p_a} \p _{x_1} \ln E(p_a,x_1) \varpi ^* _a(x_2)
+ \p _{p_a} \p _{x_2} \ln E(p_a,x_2) \varpi ^* _a(x_1) \biggr ]
\nonumber \\
\X _5 &=&
{\zeta ^1 \zeta ^2 \over 16 \pi ^2}
\sum _a \biggl [ 
S_\delta (p_a, x_1) \p _{p_a} S_\delta (p_a,x_2) 
- S_\delta (p_a, x_2) \p _{p_a} S_\delta (p_a,x_1) \biggr ] 
\varpi _a  (x_1,x_2) \, .
\nonumber
\eea
Each expression is perfectly well-defined and each term is finite for
generic points.

\subsection{The $\delta$-function gravitino expression for $\X_6$}

In order to set $\chi _\alpha (w) = \delta (w,x_\alpha)$ in $\X_6$,
we have to proceed with some extra care, since 
a singularity seems to emerge of the form $G_{3/2}(x_\alpha, x_\alpha)$
multiplying an expression trilinear in $\chi$. 
This singularity is however only apparent,
since it is naturally cancelled 
by the symmetry properties amongst the
gauge slice functions $\chi _1$ and $\chi_2$ entering into this trilinear
expression. The limit may then be taken safely and a good expression for
$\X_6$ obtained. We present now a detailed account of
this symmetrization and limiting process.

\medskip

We begin by {\sl keeping the slice functions $\chi _\alpha$ arbitrary and
regular}, letting them tend to $\delta$-functions only after all singular
contributions to $\X_6$ have cancelled out, and the limit can  be taken
safely. We need the following useful identity,
\be
\label{rearrangementchi}
\sum _\alpha \chi _\alpha ^* (z) \psi ^* _\alpha (v)
=
\sum _\alpha \chi _\alpha (z) \bar \psi _\alpha (v)\, .
\ee
Here, $ \psi ^* _\alpha (v)$ are the holomorphic 3/2 differentials with
normalization $\psi ^* _\alpha (q_\beta ) = \delta _{\alpha \beta}$, and
$\chi ^* _\beta$ are the linear combinations of $\chi_\beta$ dual to
$\psi ^* _\alpha$, so that $\< \chi ^* _\alpha | \psi ^* _\beta \> =
\delta _{\alpha \beta}$. The holomorphic 3/2 differentials $\bar \psi
_\alpha$ are then defined by (\ref{rearrangementchi}), which implies
$\bar \psi _\alpha (z) \< \chi _\alpha | \psi ^* _\beta \> = \psi ^*
_\beta (z)$, and this equation may be solved by 
\bea
\label{barpsi}
\bar \psi _1 (v) &=&
{\psi ^* _1 (v) \< \chi _2 | \psi ^* _2 \> - 
 \psi ^* _2 (v) \< \chi _2 | \psi ^* _1 \>
\over 
\< \chi _1 | \psi ^* _1 \> \< \chi _2 | \psi ^* _2 \> - 
\< \chi _2 | \psi ^* _1 \> \< \chi _1 | \psi ^* _2 \> }
\nonumber \\
\bar \psi _2 (v) &=&
{\psi ^* _2 (v) \< \chi _1 | \psi ^* _1 \>  - 
 \psi ^* _1 (v) \< \chi _1 | \psi ^* _2 \>
\over 
\< \chi _1 | \psi ^* _1 \> \< \chi _2 | \psi ^* _2 \> - 
\< \chi _2 | \psi ^* _1 \> \< \chi _1 | \psi ^* _2 \> }
\eea
As $\chi _\alpha (z) \to \delta (z,x_\alpha)$, $\bar \psi _\alpha$ has a
smooth limit, the result becomes independent of the points $q_\alpha$ and
normalized by $\bar \psi _\alpha (x_\beta ) = \delta _{\alpha \beta}$. To
evaluate the term ${\cal X}_6$, we first write all contributions in terms
of $\chi _\alpha$ instead of $\chi ^* _\alpha $, using the above formula
(\ref{rearrangementchi}). 
\bea
\label{Xsix}
\X_6 &=&
{\zeta ^1 \zeta ^2 \over 16 \pi ^2} 
\int \! d^2 \! z \int \! d^2 \! w \int \! d^2 \! v
\biggl [ 
+\chi _1(z) \chi _1(w) \chi _2(v) G_{3/2} (z,w) \bar \Lambda _1 (w,v)
\nonumber \\
&& \hskip 1.65in  
- \chi _1(z) \chi _2(w) \chi _1(v) G_{3/2} (z,w) \bar \Lambda _1 (w,v)
\nonumber \\
&& \hskip 1.65in  
+ \chi _2(z) \chi _1(w) \chi _2(v) G_{3/2} (z,w) \bar \Lambda _2 (w,v)
\nonumber \\
&& \hskip 1.65in  
- \chi _2(z) \chi _2(w) \chi _1(v) G_{3/2} (z,w) \bar \Lambda _2 (w,v)
\biggr ]
\eea
where $\bar \Lambda _\alpha$ is obtained by replacing $\psi ^* _\alpha$
by $\bar \psi _\alpha$ in $\Lambda _\alpha$ of (\ref{Lambda}),
\be
\bar \Lambda _\alpha (w,v) 
=
2 G_2(w,v) \p _v \bar \psi _\alpha (v) 
+ 3 \p _v G_2 (w,v) \bar \psi _\alpha (v)
\ee
The first term in (\ref{Xsix}) appears to generate a singularity
$G_{3/2}(x_1, x_1)$ as $\chi _1 (z) \to \delta (z,x_1)$. However, the
simple pole of $G_{3/2}(z,w)$ is odd under the interchange of $z$ and
$w$, while the product $\chi _1(z) \chi _1 (w)$ is even under this
exchange. Thus, {\sl for any regular $\chi _1$, the pole term cancels} and
the limit $\chi _1 (z) \to \delta (z,x_1)$ is regular and may now be
taken safely.

\medskip

We begin by carrying out the symmetrization explicitly~: in $z$ and $w$
in the first and fourth terms; in $z$ and $v$ in the second and third
terms. Regrouping terms, we find 
\bea
\X_6 &=&
{\zeta ^1 \zeta ^2 \over 16 \pi ^2} 
\int \! d^2 \! z \int \! d^2 \! w \int \! d^2 \! v
\biggl [ 
\\ 
&& 
\hskip .5in
+\half \chi _1(z) \chi _1(w) \chi _2(v) \biggl \{ 
G_{3/2} (z,w) \bar \Lambda _1 (w,v) + G_{3/2} (w,z) \bar \Lambda _1 (z,v)
\nonumber \\
&& \hskip 1.4in 
- G_{3/2} (z,v) \bar \Lambda _1 (v,w) - G_{3/2} (w,v) \bar \Lambda _1 (v,z)
\biggr \}
\nonumber \\
&& 
\hskip .5in 
+ \half \chi _2(z) \chi _1(w) \chi _2(v) \biggr \{
 G_{3/2} (z,w) \bar \Lambda _2 (w,v) + G_{3/2} (v,w) \bar \Lambda _2 (w,z)
\nonumber \\
&& \hskip 1.4in  
- G_{3/2} (z,v) \bar \Lambda _2 (v,w) - G_{3/2} (v,z) \bar \Lambda _2 (z,w)
\biggr \}
\biggr ]\, .
\nonumber 
\eea
Now we are ready to take the limit in which $\chi _\alpha (z) \to \delta
(z,x_\alpha)$. In this limit, $w\to z$ in the first braces, while $v\to z$ in
the second braces above. The terms that do not manifestly admit a limit 
may be evaluated with the help of the asymptotics of the Green function,
\be
G_{3/2}(x,y) = {1 \over x-y} + f_{3/2} (x) + \O (x-y)
\ee
so that
\bea
&&
\lim _{w \to z} \biggl \{+ G_{3/2} (z,w) \bar \Lambda _1 (w,v) 
+ G_{3/2} (w,z) \bar \Lambda _1 (z,v) \biggr \}
\nonumber \\ && \hskip 1.5in
= - \pz \bar \Lambda _1 (z,v) + 2 f_{3/2} (z) \bar \Lambda _1 (z,v)
\nonumber \\
&&
\lim _{v\to z} \biggl \{ -G_{3/2} (z,v) \bar \Lambda _2 (v,w) 
- G_{3/2} (v,z) \bar \Lambda _2 (z,w) \biggr \}
\nonumber \\ && \hskip 1.5in
= + \pz \bar \Lambda _2 (z,w) - 2 f_{3/2} (z) \bar \Lambda _2 (z,w)
\nonumber 
\eea
Next, we evaluate $\bar \Lambda _\alpha $ and the derivatives of $\bar
\Lambda _\alpha$ needed in the above expressions as follows
\bea
\bar \Lambda _1 (x_1, x_2) 
  &=& 2 G_2 (x_1,x_2) \p \bar \psi _1 (x_2) 
\nonumber \\
\bar \Lambda _2 (x_2, x_1) 
  &=& 2 G_2 (x_2,x_1) \p \bar \psi _2 (x_1) 
\nonumber \\
\p _{x_1} \bar \Lambda _1 (x_1, x_2) 
  &=& 2 \p _{x_1} G_2 (x_1,x_2) \p \bar \psi _1 (x_2) 
\nonumber \\
\p _{x_2} \bar \Lambda _2 (x_2, x_1) 
  &=& 2 \p _{x_2} G_2 (x_2,x_1) \p \bar \psi _2 (x_1) 
\nonumber \\
\bar \Lambda _1 (x_2, x_1) 
  &=& 2 G_2 (x_2,x_1) \p \bar \psi _1 (x_1) + 3 \p _{x_1} G_2 (x_2,x_1)
\nonumber \\
\bar \Lambda _2 (x_1, x_2) 
  &=& 2 G_2 (x_1,x_2) \p \bar \psi _2 (x_2) + 3 \p _{x_2} G_2 (x_1,x_2)
\eea
The quantities $\bar \psi_\alpha$ simplify in the limit $\chi_\alpha (z)
\to \delta (z,x_\alpha)$ of (\ref{barpsi}) and the simplified expressions
are given by
\bea
\label{barpsiexplicit}
\bar \psi _1 (v) &=&
{\psi ^* _1 (v)  \psi ^* _2 (x_2)  - \psi ^* _2 (v) \psi ^* _1 (x_2)
\over 
\psi ^* _1 (x_1) \psi ^* _2 (x_2)  - \psi ^* _1 (x_2) \psi ^* _2 (x_1) }
\nonumber \\
\bar \psi _2 (v) &=&
{\psi ^* _2 (v) \psi ^* _1 (x_1)   -  \psi ^* _1 (v) \psi ^* _2 (x_1)
\over 
\psi ^* _1 (x_1) \psi ^* _2 (x_2)  - \psi ^* _1 (x_2) \psi ^* _2 (x_1) }
\eea
 We may now assemble all contributions into a
final expression for $\X_6$,
\bea
\label{x6}
\X_6 &=&
{\zeta ^1 \zeta ^2 \over 16 \pi ^2} 
\biggl [
3 G_{3/2} (x_2,x_1) \p _{x_2} G_2(x_1,x_2) 
    - 3 G_{3/2}(x_1,x_2) \p _{x_1} G_2 (x_2,x_1) 
\nonumber \\
&& \qquad \ 
+ 2 G_{3/2} (x_2,x_1)  G_2(x_1,x_2) \p \bar \psi _2(x_2) 
    - 2 G_{3/2}(x_1,x_2)  G_2 (x_2,x_1) \p \bar \psi _1 (x_1)
\nonumber \\
&& \qquad \ 
+ 2 f_{3/2} (x_1) G_2(x_1,x_2) \p \bar \psi _1 (x_2)
    -2 f_{3/2} (x_2) G_2 (x_2,x_1) \p \bar \psi _2 (x_1)
\nonumber \\
&& \qquad \
+ \p _{x_2} G_2(x_2,x_1) \p \bar \psi _2 (x_1)
    - \p _{x_1} G_2(x_1,x_2) \p \bar \psi _1 (x_2) \biggr ]
\eea
which is perfectly well-defined and finite.

\medskip

It is worth pointing out that the sum $\X_1 +\X_6$ exhibits considerable
simplification, as the terms multiplied by 3 occurring in $\X_6$ cancel
those occurring in $\X_1$, and the total gives,
\bea 
\label{XoneplusXsix}
\X_1 + \X _6 &=&
{\zeta ^1 \zeta ^2 \over 16 \pi ^2}
\biggl [
-10 S_\delta (x_1,x_2) \p _{x_1} \p _{x_2} \ln E(x_1,x_2)
\\
&&
\qquad \
 - 2 G_2 (x_1,x_2) \p _{x_2} G_{3/2}(x_2,x_1) 
 + 2 G_2 (x_2,x_1) \p _{x_1} G_{3/2}(x_1,x_2)
\nonumber \\
&&
\qquad \
+ 2 G_{3/2} (x_2,x_1)  G_2(x_1,x_2) \p \bar \psi _2(x_2) 
    - 2 G_{3/2}(x_1,x_2)  G_2 (x_2,x_1) \p \bar \psi _1 (x_1)
\nonumber \\
&& \qquad \ 
+ 2 f_{3/2} (x_1) G_2(x_1,x_2) \p \bar \psi _1 (x_2)
    -2 f_{3/2} (x_2) G_2 (x_2,x_1) \p \bar \psi _2 (x_1)
\nonumber \\
&& \qquad \
+ \p _{x_2} G_2(x_2,x_1) \p \bar \psi _2 (x_1)
    - \p _{x_1} G_2(x_1,x_2) \p \bar \psi _1 (x_2) \biggr ]
\nonumber 
\eea
Together with the results of (\ref{regularX}), the above formula yields
the chiral superstring measure evaluated on $\delta$-function supported
gravitino slices.

\vfill\eject

\section{Global Slice $\chi$ Independence}
\setcounter{equation}{0}

We shall now prove that the full chiral superstring measure 
$\A[\delta]$, given by (\ref{Zformula}), (\ref{regularX}) and (\ref{x6}),
is a holomorphic scalar function in $x_\alpha$, $q_\alpha$ and $p_a$ by
showing that no singularities occur when any of these points pairwise
coincide. Since the measure is a holomorphic scalar in $x_1$, for
example, it must be independent of $x_1$.  By iterating this argument for
all points, we establish that $\A [\delta]$ is independent of all points
$x_\alpha$, $q_\alpha$ and $p_a$. Thus, $\A[\delta]$ is globally
independent of the choice of $\delta$-function slices.
We present below the arguments for the absence of singularities when
points coincide in order of increasing difficulty.

\subsection{Regularity as $q_\alpha \to p_a$}

This is the easiest case, as the overall factor $\Z$ as well as each term
$\X_i$, $i=1,\cdots ,6$ have a finite limit. This is manifest for
all $\X_i$, except perhaps $\X_3$, where the result follows, however, from
holomorphicity in $w$ of the functions $B_2(w)$ and $B_{3/2}(w)$.

\subsection{Regularity as $q_2 \to q_1$}

The overall factor $\Z$
\bea
\Z = {\< \prod _a b(p_a) \prod _\alpha \delta (\beta (q_\alpha)) \>
\over \det \omega _I \omega _J (p_a) 
 \cdot \det  \psi ^* _\beta (x_\alpha)}
\eea
has $q_\alpha$-dependence through both the correlator and the finite
dimensional determinant $\det  \psi ^* _\beta (x_\alpha)$. The
$q$-dependence of the latter may be exhibited using {\sl any
$q_\alpha$-independent basis of 3/2 holomorphic differentials} $\psi _1$,
$\psi _2$. We then have
\bea
\det  \psi ^* _\beta (x_\alpha)
=
{\psi _1 (x_1) \psi _2 (x_2) - \psi _1 (x_2) \psi _2 (x_1)
\over
\psi _1 (q_1) \psi _2 (q_2) - \psi _1 (q_2) \psi _2 (q_1)}
\eea
The $q_\alpha$-dependence of the correlator may also be rendered
completely explicit,
\bea
\Z = {\tet [\delta ](0)^5 \tet (D_b) \prod _{a<b} E(p_a,p_b) \prod _a
\sigma (p_a)^3
\over Z^{15} \tet [\delta ](q_1+q_2-2 \Delta) E(q_1,q_2) \sigma (q_1)^2
\sigma (q_2)^2 \ \det \omega _I \omega _J (p_a) \cdot \det  \psi ^*
_\beta (x_\alpha)}
\eea
The numerator of $\Z$ is $q_\alpha$-independent. The denominator has a
simple pole as $q_2 \to q_1$ from the factor $\det \psi ^* _\beta
(x_\alpha)$, and this pole is cancelled by a simple zero from the prime
form $E(q_1,q_2)$, leaving a finite limit of $\Z$. The
Green's functions $G_{3/2}(z,w)$ and
$f_{3/2}(z)$ have a smooth limits as $q_2 \to q_1$, and these limits are
given by (see Appendix A of \cite{II})
\bea
G_{3/2}( z,w) & = & {\tet [\delta ] (z-w + 2 q_1 - 2 \Delta) \over \tet
[\delta ] (2 q_1 - 2 \Delta) E(z,w) }
{E(z,q_1)^2 \sigma (z)^2  \over E(w,q_1)^2 \sigma (w)^2}
\nonumber \\
f_{3/2}(z) & = & \omega _I (z) \p _I \ln \tet [\delta ](2q_1 - 2 \Delta)
+ \p_z \ln \biggl ( \sigma (z)^2 E(z,q_1)^2 \biggr )
\eea
As a result, $\X_1 + \X_6$ has a smooth limit, as do $\X_2$ and $\X_3$.
The terms $\X_4$ and $\X_5$ are independent of $q_\alpha$ altogether, so
their limit is manifestly smooth.

\subsection{Regularity as $x_\alpha \to q_\alpha$}

The limit $x_\alpha \to q_\alpha$ is not well-defined term by term,
beginning with the superghost correlator $\X_1$. Here, we show that the
combination of all contributions in (\ref{Zformula}), (\ref{regularX})
and (\ref{x6}) is well-defined and finite. To begin with, it is manifest
that the prefactor $\Z$ has a well-defined limit, with the only
$x_\alpha$ dependence through $\det\psi ^* _\beta (x_\alpha) \to 1$ in
this limit. Furthermore, the terms ${\cal X}_2$, ${\cal X}_3$,
${\cal X}_4$ and ${\cal X}_5$ all have smooth limits. Thus, only the term
${\cal X}_1 + {\cal X}_6 $ remains to be examined, which we do next.

\medskip

The holomorphic differentials $\bar \psi _\alpha (z)$ behave smoothly as
$x_\beta \to q_\beta$, as do their derivatives. Thus the first and last
lines  in (\ref{XoneplusXsix}) admit smooth limits, and only
terms of the following form remain to be discussed,
\be
G_2(x_1,x_2) \biggl [-2 \p _{x_2} G_{3/2}(x_2,x_1) + 2 G_{3/2} (x_2,x_1) 
\p \bar \psi _2 (x_2) + 2 f_{3/2}(x_1) \p \bar \psi _1(x_2) \biggr ]
\ee
(minus the same form with $x_1$ and $x_2$  as well as $\bar \psi _1$ and
$\bar \psi _2$ interchanged). The above contribution exhibits a
singularity in the form of a simple pole in $x_1-q_1$, but is
regular as $x_2 \to q_2$. The pole is easily evaluated using te following
formulas
\bea
G_{3/2}(x_2,x_1) &=& { 1 \over x_1 - q_1} \psi ^* (x_2) + \O(1) 
\nonumber  \\ 
f_{3/2}(x_1) &=& {1 \over x_1 - q_1} + \O(1)
\eea
The residue of the pole is given by
\be
-2 \p \psi ^*_1 (x_2) + 2 \psi ^* _1 (x_2) \p \bar \psi _2 (x_2) + 2 \p 
\bar
\psi _1 (x_2)\, ,
\ee
a formula in which $x_1 = q_1$ since we are evaluating the residue at the
pole in $(x_1-q_1)$. With this value for $x_1$, the $\bar  \psi$
differentials (\ref{barpsi}) simplify considerably and we have
\bea
\bar \psi _1 (x) = \psi ^* _1 (x) - \psi ^* _2(x) {\psi ^* _1(x_2) \over
\psi ^* _2(x_2)}
\hskip 1in
\bar \psi _2 (x) = {\psi ^* _2 (x) \over \psi ^* _2 (x_2) }
\eea
Using these expressions, the residue is readily seen to vanish. We
conclude that the limit $x_\alpha \to q_\alpha$ is smooth in the full 
chiral superstring measure.

\subsection{Regularity as $p_a \to p_b$}

The Green's function $G_2$ behaves smoothly in this limit, while
$S_\delta$, $G_{3/2}$ and $f_{3/2}$ are simply independent of $p_a$. As a
result, $\X_1 + \X_6$ and $\X_2$ have smooth limits as two $p_a$ collapse.
The limits of $\X_3$, $\X_4$ and $\X_5$ are more involved as the forms
$\varpi ^*_a$ and $\varpi _a$ have implicit $p_a$ dependence which
may become singular. To study this behavior, we fix $p_1 \not= p_2$ and
let $p_3 \to p_1$ without loss of generality. The terms $\X_3$, $\X_4$
and $\X_5$ are now all of the form
\bea
\sum _a \varpi ^*_a (x) f(p_a)
\hskip 1in 
\sum _a \varpi  _a (x_1,x_2) f(p_a)
\eea
and we shall show that this limit is smooth provided $f$ is
differentiable, which is of course the case here. To analyze $\varpi ^*_a$
and $\varpi _a$ in this configuration, it is convenient to choose an
adapted basis for holomorphic Abelian differentials, $\omega ^* _I(p_J) =
\delta _{IJ}$ for $I,J=1,2$. We then have considerably simplified and
more workable expressions for $\varpi ^*_a$ and $\varpi _a$, given by
\bea
\varpi ^* _1 (x) & = & \omega ^* _1 (x) - \half {\omega ^* _1 (p_3) \over
\omega ^* _2 (p_3)} \omega ^* _2 (x)
\nonumber \\
\varpi ^* _2 (x) & = & \omega ^* _2 (x) - \half {\omega ^* _2 (p_3) \over
\omega ^* _1 (p_3)} \omega ^* _1 (x)
\nonumber \\
\varpi ^* _3 (x) & = & \half \biggl ( {\omega ^*_1 (x) \over \omega ^*
_1(p_3) } + {\omega ^*_2 (x) \over \omega ^* _2 (p_3) } \biggr )
\eea
and
\bea
\varpi  _1 (x_1,x_2) & = & 
\omega ^* _1 (x_1) \omega ^* _1 (x_2)
- \half \omega ^* _1 (p_3) {\omega ^* _1 (x_1) \omega ^* _2 (x_2) + \omega
^* _1 (x_2) \omega ^* _2 (x_1) \over \omega ^* _2 (p_3)} 
\nonumber \\
\varpi  _2 (x_1,x_2) & = & 
\omega ^* _2 (x_1) \omega ^* _2 (x_2)
- \half \omega ^* _2 (p_3) {\omega ^* _1 (x_1) \omega ^* _2 (x_2) + \omega
^* _1 (x_2) \omega ^* _2 (x_1) \over \omega ^* _1 (p_3)} 
\nonumber \\
\varpi  _3 (x_1,x_2) & = & 
\half {\omega ^* _1 (x_1) \omega ^* _2 (x_2) + \omega ^* _1 (x_2) \omega
^* _2 (x_1) \over \omega ^* _1 (p_3) \omega ^* _2 (p_3)} 
\eea
The limits as $p_3 \to p_1$ of the sums are now easily evaluated and we
find
\bea
\sum _a \varpi ^* _a (x) f(p_a)
& = &
{3 \over 2} \omega _1 ^* (x) f(p_1) + \omega ^* _2 (x) f(p_2)
+ \half {\omega ^* _2 (x) \p f(p_1) \over \p \omega ^* _2(p_1) } 
\\
\sum _a \varpi  _a (x_1, x_2) f(p_a)
& = &
\omega ^* _1 (x_1) \omega ^* _1 (x_2) f(p_1) + \omega ^* _2 (x_1) \omega
^* _2 (x_2) f(p_2) 
\nonumber \\ &&
+ \half \biggl ( \omega ^* _1 (x_1) \omega ^* _2 (x_2)
+ \omega ^* _1 (x_2) \omega ^* _2 (x_1) \biggr ) {\p f (p_1) \over \p
\omega _2 ^* (p_1)}
\eea
both of which are finite. This establishes that the limits of collapsing
$p_a$'s are smooth.

\subsection{Regularity as $x_2 \to x_1$}

The prefactor $\Z$ in (\ref{Zformula}) exhibits an overall  simple pole as
$x_2 \to x_1$ since the finite dimensional determinant $\det \psi ^*
_\beta (x_\alpha)$ has a simple zero in this limit. Amongst the
$\X_i$ of  (\ref{regularX}), $\X_1$ exhibits a simple pole, which is
cancelled by the simple poles in $\X_2$ and those parts of the simple
pole in $\X_3$ that are produced by the full stress tensor. The remaining
parts of $\X_3$ as well as $\X_4$ exhibit a simple pole, while $\X_5$
admits a  vanishing limit.\footnote{It is helpful to notice that the
prefactor $\Z$ as well as each $\X_i$ is odd under interchange of $x_1$
and $x_2$.}

\medskip

Only $\X_6$ appears to produce a triple pole, and we shall begin by
showing that this pole cancels within $\X_6$. The starting point is the
expression (\ref{x6}) and the limiting behaviors of $\p \bar \psi _\alpha
(x_\beta)$, given by
\bea
\p \bar \psi _1 (x_1), \ \p \bar \psi _1 (x_2) &\to & {1 \over x_1 - x_2}
\nonumber \\ 
\p \bar \psi _2 (x_1), \ \p \bar \psi _2 (x_2) &\to & {-1 \over x_1 - x_2}
\eea
which upon substitution into (\ref{x6}) leads to the absence of
the triple pole in $\X_6$.

\subsubsection{Cancellation of the simple poles in $\X_i$}

There are neither double poles nor constant terms in $\X_i$ since it is
odd under $x_1 \leftrightarrow x_2$. Therefore, it remains to show that
the simple poles cancel. To this end, an extra careful asymptotic
analysis is required. We begin by defining the variables in which the
limit will be taken~:
\be
x_1 = x+\epsilon \qquad \qquad x_2 = x-\epsilon
\qquad \qquad \epsilon \to 0 \ {\rm with} \ x \ {\rm fixed}\, .
\ee
The derivatives $\p \bar \psi _\alpha (x_\beta)$ behave as follows
\bea
\p \bar \psi _1 (x _1 ) & = & 
   +{1 \over 2 \epsilon} + \half A + {\epsilon \over 2} B +\O(\epsilon^2)
\nonumber \\
\p \bar \psi _1 (x _2 ) & = & 
   +{1 \over 2 \epsilon} - \half A + {\epsilon \over 4} (C-B)
+\O(\epsilon^2)
\nonumber \\
\p \bar \psi _2 (x _1 ) & = & 
   -{1 \over 2 \epsilon} - \half A + {\epsilon \over 4} (C-B)
+\O(\epsilon^2)
\nonumber \\
\p \bar \psi _2 (x _2 ) & = & 
   -{1 \over 2 \epsilon} + \half A - {\epsilon \over 2} B
+\O(\epsilon^2)
\eea
where $A,B,C$ are defined by the following expressions,
evaluated at $x$,
\bea
A & = & {
\psi ^* _2 \p ^2 \psi ^* _1 - \psi ^* _1 \p ^2 \psi ^* _2
\over 
\psi ^* _2 \p  \psi ^* _1 - \psi ^* _1 \p  \psi ^* _2} 
\nonumber \\
B & = & {
\third \psi ^* _2 \p ^3 \psi ^* _1 - \third \psi ^* _1 \p ^3 \psi ^* _2
+ \p \psi ^* _2 \p ^2 \psi ^* _1 - \p \psi ^* _1 \p ^2 \psi ^* _2
\over 
\psi ^* _2 \p  \psi ^* _1 - \psi ^* _1 \p  \psi ^* _2}
\nonumber \\
C & = & {
 \psi ^* _2 \p ^3 \psi ^* _1 - \psi ^* _1 \p ^3 \psi ^* _2
+ \p \psi ^* _2 \p ^2 \psi ^* _1 - \p \psi ^* _1 \p ^2 \psi ^* _2
\over 
\psi ^* _2 \p  \psi ^* _1 - \psi ^* _1 \p  \psi ^* _2}\, .
\eea
We shall also need the asymptotics of the Green's functions, which are 
given  as follows (see Appendix A of \cite{II} for more details)
\bea
\label{limitsgreen}
\p _{x_1} \p _{x_2} \ln E(x_1,x_2) &=&
+{1 \over 4 \epsilon ^2} - 2 T_1(x) +\O(\epsilon ^2)
\nonumber \\
S_\delta (x_1,x_2) &=&
+ {1 \over 2 \epsilon } + 2 \epsilon g_{1/2} (x) - 2 \epsilon T_1(x) 
+\O(\epsilon ^3 )
\nonumber \\ 
G _n (x_1 , x_2 ) &=& 
  +{1 \over 2 \epsilon} + f_n (x) + \epsilon (2g_n(x) - \p f_n (x) 
-2T_1(x)) +\O(\epsilon^2)
\nonumber \\   
G _n (x_2 , x_1 ) &=& 
  -{1 \over 2 \epsilon} + f_n (x) - \epsilon (2g_n(x) - \p f_n (x) 
-2T_1(x)) +\O(\epsilon^2)
\nonumber \\
\p _{x_1} G _n (x_1 , x_2 ) &=& 
  -{1 \over 4 \epsilon ^2} + g_n(x)  -T_1(x) +\O(\epsilon)
\nonumber \\   
\p _{x_2} G _n (x_2 , x_1 ) &=& 
  -{1 \over 4 \epsilon ^2 } + g_n(x)  - T_1(x) +\O(\epsilon)
\eea
where $f_n(w)$ is given by the first line of (\ref{effs}) and $g_n(w)$ by 
\bea
\label{geen}
g_n(w) = \half \omega _I \omega _J(w) \p _I \p_J \ln \tet [\delta ] (D_n)
+ \half f_n(w)^2  + \half \p_w f_n (w) \, .
\eea
We now calculate the limiting pole behavior of each of the terms $\X_i$.
We omit an overall factor of $\zeta ^1 \zeta ^2 / 16\pi ^2 \epsilon$
which is common to all terms. The details of the calculation of $\X_3$
will be given below.
\bea
\label{xxlimits}
\X _1 + \X_6 & \sim &
3g_2(x) +12 \ T_1(x) -2 \p f_2(x) -A f_2(x) + {1 \over 8} (3B+C) -5 
g_{1/2}(x)
\nonumber \\
\X _2 & \sim &
\half \omega \omega _J(x) \biggl [ \p _I \p _J \ln {\tet [\delta ](0)^5 
\over
\tet [\delta ](D_\beta )} +   \p _I \p _J \ln \tet (D_b ) \biggr ]
\nonumber \\
\X _3 & \sim &
-12 \ T_1(x) + \half f_2 (x)^2 - {3 \over 2} \p f_2(x) 
- 2 \sum _a \p _{p_a} \p _x \ln E (p_a,x) \varpi ^* _a(x)
- \half f_{3/2}(x)^2
\nonumber \\
&&   -\half \p f_{3/2}(x) + g_{3/2}(x) + 4 \p
f_2(x) - 4 g_2(x) + Af_2(x) - {1 \over 8}(3B+C)
\nonumber \\
\X _4 & \sim &
2 \sum _a \p _{p_a} \p _x \ln E (p_a,x) \varpi ^* _a(x)
\nonumber \\
\X _5 & \sim & 0
\eea
Adding all terms but $\X_2$, and using the expressions for $f_n$ of
(\ref{effs}) and  $g_n$ of (\ref{geen}), we find
\bea
&& \half f_2(x)^2 - \half f_{3/2}(x)^2 + \half \p f_2(x) - \half \p 
f_{3/2}(x) + g_{3/2}(x) - g_2 (x) - 5 g_{1/2} (x)
\nonumber \\
&& =
\half \omega _I \omega _J(x) \biggl [
+ \p _I \p _J \ln \tet [\delta ] (D_\beta )
- \p _I \p _J \ln \tet  (D_b )
-5\p _I \p _J \ln \tet [\delta ] (0 ) \biggr ]
\eea
and this term is readily seen to cancel completely with $\X_2$.

\subsubsection{Detailed evaluation of the limit of $\X_3$}

The one piece of the above calculation that requires further detailing is 
the evaluation of $\X_3$. As $x_1 = x+\epsilon$ and $x_2 = x-\epsilon$,
with $x$ held fixed and $\epsilon \to 0$, the pole in $\X_3$ takes the
following form
\be
{\rm pole} \ \X _3 
=
{\zeta ^1 \zeta ^2 \over 16\pi ^2} {1 \over \epsilon}
\sum _a \varpi  _a (x,x) \biggl ( B_2(p_a) + B_{3/2}(p_a) \biggr )\, .
\ee
First, we use the fact that $\varpi _a(x,x) = \phi ^{(2)*} _a(x)$, 
and then we use the fact that since $B_2$ and $B_{3/2}$ are holomorphic
2-forms,  we have 
\be 
\sum _a \phi ^{(2)*} _a (x) \biggl ( B_2 (p_a) + B_{3/2}(p_a) \biggr )
= B_2 (x) + B_{3/2}(x)\, .
\ee
The evaluation of $B_2(x)$ is straightforward. To evaluate $B_{3/2}(x)$, 
we keep $\epsilon \not= 0$, and work out its expression starting from its
definition in (\ref{bthreehalfs}). We find
\bea
\label{bx}
B_{3/2}(x) &=&
12 T_1 (x) - \half f_{3/2}(x)^2 + \p f_{3/2}(x)
\nonumber \\
&& 
- {3 \over 2} \p _x G_{3/2}(x_1,x) \bar \psi _1 (x)
- {3 \over 2} \p _x G_{3/2}(x_2,x) \bar \psi _2 (x)
\nonumber \\
&&
- {1 \over 2}  G_{3/2}(x_1,x) \p \bar \psi _1 (x)
- {1 \over 2}  G_{3/2}(x_2,x) \p \bar \psi _2 (x)
\nonumber \\
&&
+ G_2 (x,x_1) \p \bar \psi _1 (x_1) + G_2 (x,x_2) \p \bar \psi _2 (x_2)
\nonumber \\
&&
+ {3 \over 2} \p _{x_1} G_2 (x,x_1) + {3 \over 2} \p _{x_2} G_2 (x,x_2) 
\eea
To evaluate this quantity, we need further asymptotics of the Green's 
functions,\footnote{Notice that these limits are slightly different from
those of (\ref{limitsgreen}) since here one of the arguments of the Green's
functions is $x$ in contrast with (\ref{limitsgreen}), where both
arguments are $x_\alpha$. As a result, the coefficients on the right are
slightly different.}
\bea
G_{3/2} (x_1,x) 
  &=& + {1\over \epsilon} + f_{3/2}(x) + \epsilon g_{3/2}(x) -  \epsilon
T_1(x) +\O(\epsilon^2)
\nonumber \\
G_{3/2} (x_2,x) 
  &=& - {1\over \epsilon} + f_{3/2}(x) - \epsilon g_{3/2}(x) +  \epsilon
T_1(x) +\O(\epsilon^2)
\nonumber \\
\p _x G_{3/2} (x_1,x) 
  &=& + {1\over \epsilon ^2} + \p f_{3/2}(x) - g_{3/2}(x) +T_1(x)
+\O(\epsilon)
\nonumber \\
\p _x G_{3/2} (x_2,x) 
  &=& + {1\over \epsilon ^2} + \p f_{3/2}(x) - g_{3/2}(x) +T_1(x)
+\O(\epsilon)
\eea
and 
\bea
G_2 (x,x_1) 
  &=& - {1\over \epsilon} + f_2(x) + \epsilon \p f_2 (x) - \epsilon g_2(x)
      + \epsilon T_1(x) +\O(\epsilon^2)
\nonumber \\
G_2 (x,x_2) 
  &=& + {1\over \epsilon} + f_2(x) - \epsilon \p f_2 (x) + \epsilon g_2(x)
      - \epsilon T_1(x) +\O(\epsilon^2)
\nonumber \\
\p _{x_1} G_2 (x,x_1) 
  &=& + {1\over \epsilon ^2} + \p f_2(x) - g_2(x) +T_1(x) +\O(\epsilon)
\nonumber \\
\p _{x_2} G_2 (x,x_2) 
  &=& + {1\over \epsilon ^2} + \p f_2(x) - g_2(x) +T_1(x)+\O(\epsilon)\, .
\eea
We also need new asymptotics of $\bar \psi _\alpha $ and its derivatives
\bea
\bar \psi _1 (x) 
  &=& +\half - {\epsilon \over 4} A + {\epsilon ^2 \over 16} (-3B+C)
+\O(\epsilon^3)
\nonumber \\
\bar \psi _2 (x) 
  &=& +\half + {\epsilon \over 4} A + {\epsilon ^2 \over 16} (-3B+C)
+\O(\epsilon^3)
\nonumber \\
\p  \bar \psi _1 (x) 
  &=& +{1 \over 2 \epsilon} - {\epsilon \over 16} (B+C) +\O(\epsilon^2)
\nonumber \\
\p  \bar \psi _2 (x) 
  &=& -{1 \over 2 \epsilon} + {\epsilon \over 16} (B+C) +\O(\epsilon^2) 
\, .
\eea
Assembling all these pieces into the expression (\ref{bx}), we find the
result given in (\ref{xxlimits}).

\subsection{Limits as $x_{\alpha}\to p_a$}

We start from formulas (\ref{Zformula}) and (\ref{regularX}) and evaluate
the limit $x_\alpha \to p_a$ of each of the $\X_i$, $i=1,\cdots ,6$. 
Since $x_1$ and $x_2$ play symmetrical roles, we examine only the limit
$x_1 \to p_a$, without loss of generality. The only $x_1$-dependence
of the prefactor (\ref{Zformula}) is through the finite-dimensional
determinant $\det \psi ^* _\beta (x_\alpha) $, which has a finite limit
for generic points $p_a$. Thus, we shall need only the singular terms of
the limit of $\X_i$ as $x_1 \to p_a$. We shall need the following
asymptotics
\bea
\label{limits}
G_2 (z ,x_1 ) 
& = &
{1 \over x_1 - p_a} \ \phi ^{(2)*} _a (z) + \O (1)
\nonumber \\
\p _{x_1} \p _{p_a} \ln E(x_1, p_a) 
& = &
{1 \over (x_1 - p_a)^2} + \O (1)
\nonumber \\
S_\delta (x_1 , p_a) 
& = &
{1 \over x_1 - p_a} + \O (x_1 - p_a)
\eea 
All other limits, up to regular terms, such as $\p_{p_a} S_\delta
(x_1 , p_a)$ may be deduced from the above.

\medskip

In evaluating the limits of the $\X_i$, there is an overall factor of
$\zeta ^1 \zeta ^2 /16\pi ^2$ which we shall suppress here.
The various limits are then given by\footnote{We use the notation $\p _p
X (p) \big |_{p=p_a}$ here and below whenever $X$ has implicit
dependence on $p_a$, such as is the case with $X= \varpi ^*_a (p)$ and
$X= \varpi  _a (p,x_2)$.} on 
\bea
\X _1 & \sim &
- {3 \over (x_1 - p_a)^2} \ \phi ^{(2)*} _a (x_2) G_{3/2} (p_a, x_2)
- {1 \over x_1 - p_a} \ \phi ^{(2)*} _a (x_2) \p_{p_a} G_{3/2} (p_a, x_2)
\nonumber \\
\X _2 & \sim & 0
\nonumber \\
\X _3 & \sim & - {3 \over (x_1 - p_a)^2} \ S_\delta (p_a,x_2) \varpi ^*
_a (x_2) - {3 \over x_1 - p_a} \p _p \biggl (
S_\delta (p,x_2) \varpi  _a (p, x_2) \biggr ) \bigg |_{p=p_a}
\nonumber \\ && 
+ {2 \over x_1 - p_a} \ S_\delta (p_a, x_2) \p \bar \psi _1 (x_1)
\varpi ^* _a (x_2) 
\nonumber \\
\X _4 & \sim & + {2 \over (x_1 - p_a)^2} \ S_\delta (p_a,x_2) \varpi ^*
_a (x_2) + {2 \over x_1 - p_a} \ \p_{p_a} S_\delta (p_a, x_2) \varpi ^*
_a (x_2) 
\nonumber \\
\X _5 & \sim & + {1 \over (x_1 - p_a)^2} \ S_\delta (p_a,x_2) \varpi ^*
_a (x_2) + {1 \over x_1 - p_a} \ S_\delta (p_a, x_2) \ \p_p \varpi 
_a (p,x_2)  \bigg |_{p=p_a}
\nonumber \\ && 
 - {1 \over x_1 - p_a} \ \p_{p_a} S_\delta
(p_a, x_2) \varpi ^* _a (x_2) 
\nonumber \\
\X _6 & \sim & + {3 \over (x_1 - p_a)^2} \ \phi ^{(2)*} _a (x_2) G_{3/2}
(p_a, x_2) + {3 \over x_1 - p_a} \ \phi ^{(2)*} _a (x_2) \p_{p_a} G_{3/2}
(p_a, x_2)
\nonumber \\ &&
 - {2 \over x_1 - p_a} \ \phi ^{(2)*} _a (x_2) G_{3/2}
(p_a, x_2) \p \bar \psi _1 (x_1) - {2 \over x_1 - p_a} f_{3/2} (x_2)  \phi
^{(2)*} _a (x_2) \p \bar \psi _2 (x_1)
\nonumber \\ &&
 + {1 \over x_1 - p_a} \p \phi ^{(2)*} _a (x_2) \p \bar \psi _2 (x_1)
\eea
It is easily established that the coefficients of all double poles simply
cancel one another.

\medskip

The remaining simple pole at $x_1 = p_a$ has the following residue
which, after working out the $p$-derivative in $\X_3$ and regrouping
terms, takes the form,
\bea
R_a (x_2; p,q) & = &
+ 2 \phi ^{(2)*} _a (x_2) \p _{p_a} G_{3/2} (p_a, x_2) 
- 2 \phi ^{(2)*} _a (x_2) G_{3/2} (p_a, x_2) \p \bar \psi _1 (p_a)
\nonumber \\ &&
- 2 \phi ^{(2)*} _a (x_2) f_{3/2} (x_2) \p \bar \psi _2(p_a) 
+ \p \phi ^{(2)*} _a (x_2) \p \bar \psi _2 (p_a)
\nonumber \\ &&
+ 2 \varpi ^* _a (x_2) S_\delta (p_a,x_2) \p \bar \psi _1 (p_a)
- 2 \varpi ^* _a (x_2) \p_{p_a} S_\delta (p_a,x_2) 
\nonumber \\ &&
- 2  S_\delta (p_a,x_2) \p _p \varpi  _a (p,x_2) \big |_{p=p_a}
\eea
It remain to show that this residue vanishes; the arguments are 
surprisingly involved.

\medskip

The residue $R_a$ is a differential form of weight $3/2$ in $x_2$. We
begin by showing that this form has no poles in $x_2$, and is thus a
holomorphic 3/2 form. By inspecting each of the ingredients of $R_a$, it
is clear that the only possible singularities in $x_2$ can occur at one of
the points $q_\alpha$ or $p_b$. We show that there are in fact no such
poles. Once this has been established, we shall show that the holomorphic
3/2 form $R_a$ vanishes at both points $p_b$, $b\not=a$. Since these
points were generic, they cannot be the divisor of any non-vanishing
holomorphic 3/2 form, and thus $R_a=0$.

\subsubsection{Holomorphicity of the residue}

To examine the singularity structure of $R_a$ as $x_2 \to q_\alpha$, we
need the following pole structure of each individual piece. First, $\phi
^{(2)*}(x_2)$, $\p \bar \psi _\beta (p_a)$, $\varpi ^* _a (x_2)$ and $\p
_p \varpi  _a (p,x_2) \big |_{p=p_a}$ are all regular in the limit.
On the other hand, the singular ingredients are given by
\bea
G_{3/2} (p,x_2) & = & {1 \over x_2 - q_\alpha} \psi ^* _\alpha (p) + \O(1)
\nonumber \\
f_{3/2} (x_2) & = & {1 \over x_2 - q_\alpha } + \p \psi ^* _\alpha
(q_\alpha) + \O (x_2 - q_\alpha)
\eea
Only the first three terms contribute to the pole, and the limiting
behavior is given by
\bea
\label{doubleres}
R_a = {2 \phi _a ^{(2)*} (x_2) \over x_2 - q_\alpha} \biggl (
\p \psi ^* _\alpha (p_a) - \psi ^* _\alpha (p_a) \p \bar \psi _1 (p_a) -
\p \bar \psi _2 (p_a) \biggr ) + \O (1)
\eea
The definition of $\bar \psi _\beta$ in (\ref{barpsiexplicit}) involves
$\psi ^* _\gamma$ as well as the values of $x_1$ and $x_2$. Since the
residue $R_a$ is evaluated at $x_2 = q_\alpha$, these definitions simplify
considerably and may be conveniently expressed in terms of $\psi ^*
_\gamma$. Let $\alpha =1$ without loss of generality,
\bea
\bar \psi _1 (p) = { \psi ^* _2 (p) \over \psi ^* _2 (x_1)}
\hskip 1in
\bar \psi _2 (p) = \psi ^* _1 (p) - \psi ^* _2 (p) {\psi ^* _1 (x_1)
\over \psi ^* _2 (x_1)}
\eea
for any point $p$. By differentiating in $p$ and setting $p=x_1$, we get
\bea
\p \bar \psi _1 (x_1) = { \p \psi ^* _2 (x_1) \over \psi ^* _2 (x_1)}
\hskip 1in
\p \bar \psi _2 (x_1) = \p \psi ^* _1 (x_1) - \p \psi ^* _2 (x_1) {\psi ^*
_1 (x_1) \over \psi ^* _2 (x_1)}
\eea
As a result, the factor in brackets in (\ref{doubleres}) vanishes for any
point $x_1=p_a$, $ \p \psi ^* _\alpha (x_1) - \psi ^* _\alpha (x_1) \p
\bar \psi _1 (x_1) - \p \bar \psi _2 (x_1)=0 $
thereby showing that the pole of (\ref{doubleres}) is absent as $x_2 \to
q_1$. The case $x_2 \to q_2$ is analogous. Thus, $R_a$ has no poles as
$x_2 \to q_\alpha$.

\medskip

To examine the singularity structure as $x_2 \to p_b$, we have to deal
with two distinct cases. When $b\not= a$, the first three terms in $R_a$
tend to zero as $\phi ^{(2)*}_a (p_b)=0$, while the remaining terms have
a finite limit. When $b=a$, double and single poles are generated, 
\bea
R_a \sim {4 \over (x_2 - p_a)^2} \biggl ( - \phi ^{(2)*}_a (x_2) + \varpi
^* _a (x_2) \biggr ) + 
{1 \over x_2 - p_a} \biggl ( \p _{x_2} \phi ^{(2)*}_a (x_2) + 2 \p \varpi
^* _a (p_a) \biggr )
\eea
Expanding the argument $x_2$ around $p_a$ in the double pole terms and
using the identity 
$$
\p _{x_2} \phi ^{(2)*}_a (x_2) \bigg |_{x_2 = p_a} = 2 \p
_{x_2} \varpi ^* _a (x_2 ) \bigg |_{x_2 = p_a}
$$ 
we see that this quantity cancels. Thus, the limit $x_2 \to p_b$ of $R_a$
is regular as well.

\subsubsection{Vanishing of the residue}

It remains to show that $R_a=0$ at the points $x_2 = p_b$ for $b\not=a$. 
The residue function simplifies at these values, and we have
\bea
R_a (p_b; p,q) & = &
 + \p \phi ^{(2)*} _a (p_b) \p \bar \psi _2 (p_a)
- 2  S_\delta (p_a,p_b) \p _p \varpi  _a (p,p_b) \bigg |_{p=p_a}
\nonumber \\ &&
+ 2 \varpi ^* _a (p_b) S_\delta (p_a,p_b) \p \bar \psi _1 (p_a)
- 2 \varpi ^* _a (p_b) \p_{p_a} S_\delta (p_a,p_b) 
\eea
As all points $p_a$ are on an equal footing, we may choose, without loss
of generality, $a=3$ and $b=1,2$. It suffices to show that
$R_3 (p_1;p,q)=0$; the same argument may then be applied to show that 
$R_3 (p_2;p,q)=0$ as well.

\medskip

To demonstrate that $R_3 (p_1; p,q)=0$, we evaluate $\phi
^{(2)*} _3$, $\varpi ^* _3 (p_1)$ and $\p \varpi  _3 (p_3,p_1)$ in a
common basis where their expressions may be compared. To this end, we
introduce (as in subsection \S 3.4) a basis of holomorphic Abelian
differentials $\omega _I ^* $ normalized so that $\omega ^* _I (p_J) =
\delta _{IJ}$, $I,J=1,2$. In terms of these objects, we have 
\bea
\p \phi ^{(2)*} _3 (p_1) & = & 
{\p \omega ^* _2 (p_1) \over \omega ^* (p_3) \omega ^* _2 (p_3) }
\hskip 1in
2 \p \varpi ^* _3 (p_1) = 
{ \omega ^* _2 (p_3) \over \omega ^* (p_3) \omega ^* _2 (p_3) }
\nonumber \\
2 \p_{p_3} \varpi  _3 (p_1, p_3) & = & 
{ \p \omega ^* _2 (p_3) \over \omega ^* (p_3) \omega ^* _2 (p_3) }
\eea
Using these expressions, we may recast $R_3(p_1; p,q)$ in the following
form,
\bea
\label{RRR}
R_a (p_b; p,q) & = &
{1 \over \omega ^* _2 (p_3) } \biggl (
\p \ln \omega ^* _2 (p_1) \p \psi ^* _1 (p_3) + \p \ln \omega ^* _2 (p_3)
S_\delta (p_1,p_3) 
\nonumber \\ && \hskip .6in 
+ \p _{p_3} S_\delta (p_1,p_3) - \p \psi ^* _2 (p_3) S_\delta (p_1,p_3)
\biggr )
\eea
The quantity in parentheses is in fact independent of $p_2$, as may be
established by noticing that $S_\delta$ and $\p \psi ^* _1 (p_2)$ are
independnt of $p_2$, and that the remaining quantities are given by
\bea
{\p \omega ^* _2 (p_1) \over \omega ^* _2 (p_3)} & = &
{\tet (2p_1 - w - \Delta)  \over \tet (p_1 + p_3 - w - \Delta)}
{E(p_3,w) \over E(p_1,w) E(p_3,p_1)} {\sigma (p_1) \over \sigma (p_3)}
\\
\p \ln \omega ^* _2 (p_3) & = &
\omega _I (p_3) \p _I \tet (p_1 + p_3 - w - \Delta) + \p _{p_3} \ln
\biggl ({E(p_3,p_1) \sigma (p_3) \over E(p_3, w)} \biggr )
\eea
where $w$ is an aribtrary point. Since $1/\omega ^* _2 (p_3) \not=0$,
showing the vanishing of $R_a (p_b; p,q)$ in (\ref{RRR}) is equivalent to
showing the vanishing of
$\omega ^* _2 (p_3)R_a (p_b; p,q)$, which is just the bracket in
(\ref{RRR}). This quantity is a form of weight 1/2 in $p_1$, and its only
possible singularities are when $p_1 \to p_3$. To show that this quantity
vanishes, it suffices to show that it is holomorphic, since with even spin
structure there are no holomorphic 1/2 forms. It suffices to pick up the
poles as $p_1 \to p_3$, which may be done with the help of
\bea
\p \psi ^* _1 (p_2) 
& \sim &
{1 \over p_1 - p_2} - \omega _I (p_1) \p _I \ln \tet [\delta ] (D_\beta )
 - 2 \p _{p_1} \ln \sigma (p_1)
\nonumber \\
{\p \omega ^* _2  (p_1) \over \omega _2 (p_2) }
& \sim &
- {1 \over p_1 - p_2} - \omega _I (p_1) \p _I \ln \tet (2p_1 -w_0 -\Delta)
 - \p _{p_1} \ln \sigma (p_1) + \p _{p_1} \ln E(p_1,w_0)
\nonumber \\
\p \ln \omega _2(p_2) 
& \sim &
- {1 \over p_1 - p_2} + \omega _I (p_1) \p _I \ln \tet (2p_1 -w_0 -\Delta)
 + \p _{p_1} \ln \sigma (p_1) - \p _{p_1} \ln E(p_1,w_0)
\nonumber \\
\p \psi ^* _2 (p_2) 
& \sim & 
{1 \over p_2 - p_1} + \omega _I (p_1) \p _I \ln \tet [\delta ] (D_\beta )    
+ 2 \p _{p_1} \ln \sigma (p_1)
\eea
and we see that all terms cancel. This concludes the proof of the fact
that $R_a(x_2;p,q)=0$, and thus of the fact that the limits $x_\alpha \to
p_a$ are regular.

\vfill\eject

\section{The Limit $x_\alpha \to q_\alpha$ and Picture Changing Operators}
\setcounter{equation}{0}

The expression for the chiral superstring measure now involves 7 distinct 
generic points, $x_\alpha, \ q_\alpha $ and $p_a$, upon which the actual
amplitude does not depend. Clearly, one would like to do away with any
reference to specific points in the final form of the chiral measure. One
way to proceed is to let various points come together and collapse; all
such limits are regular. Another way is to make special choices for the
points without actually collapsing them. We shall make use of both
approaches. We conclude this paper with the derivation of the final
formulas (\ref{finamppq}) and (\ref{peeque}), 
in which the points $x_\alpha$ and $q_\alpha$ have been collapsed onto
one another and $x_\alpha = q_\alpha$.  The resulting formula is the
starting point of the next paper IV in this series, where the chiral
superstring measure will be cast in terms of modular forms.

\medskip

The limit $x_\alpha \to q_\alpha$ produces very significant
simplifications such as $\det  \psi ^* _\beta (x_\alpha) =1$, while
$B_{3/2}(w)$ and some of the Green's functions simplify. The necessary
ingredients are,
\bea
G_{3/2} (x_2, x_1) &=&
{1 \over x_1 - q_1} \psi ^* _1 (x_2) 
      - \psi ^* _1 (x_2) f_{3/2} ^{(1)} (x_2) +\O (x_1 - q_1)
\nonumber \\
G_{3/2} (x_1, x_2) &=&
{1 \over x_2 - q_2} \psi ^* _2 (x_1) 
      - \psi ^* _2 (x_1) f_{3/2} ^{(2)} (x_1) +\O (x_2 - q_2)
\nonumber \\
f_{3/2} (x_1) &=&
{1 \over x_1 -q_1} + \p \psi ^* _1 (q_1) + \O (x_1 - q_1) 
\nonumber \\
f_{3/2} (x_2) &=&
{1 \over x_2 -q_2} + \p \psi ^* _2 (q_2) + \O (x_2 - q_2)\, ,
\eea 
where we use the definitions of $f_{3/2} ^{(\alpha)} (w)$, given in
(\ref{effs}).
Clearly, since the points $p_a$ have been kept separate from the points
$q_\alpha$, the limits $x_\alpha \to q_\alpha $ on the Green function
$G_2$  are regular. Similarly, the limit of the matter contribution is
regular.  The remaining contributions involve $G_{3/2} (x_1,x_2)$ and
$G_{3/2} (x_2,x_1)$ respectively, of which the second is given by
\bea
-2 \p _{x_2} G_{3/2} (x_2,x_1)
+2 G_{3/2} (x_2,x_1) \p \bar \psi _2 (x_2)
+ 2 f_{3/2}(x_1) \p \bar \psi _1 (x_2)\, .
\eea
The $x_1 \to q_1$ limit of this quantity is regular, as was already shown 
in the preceding section, as a simple pole is cancelled between the three
terms. The $x_2 \to q_2$ limit is regular for every term by itself, so
both
$x_\alpha \to q_\alpha$ limits are smooth and may be taken in any order. 
We begin by taking the limit $x_2 \to q_2$ first, which results in
\bea
\label{partq}
-2 \p _{q_2} G_{3/2} (q_2,x_1)
+ 2 f_{3/2}(x_1) \p \bar \psi _1 (q_2)\, ,
\eea
since $G_{3/2}(q_2,x_1)=0$. To take the limit $x_1 \to q_1$ next, we need 
to evaluate the factor $\p \bar \psi _1 (q_2)$, for which we use the fact
that for $x_2 = q_2$, we have $\bar \psi _1(x) = \psi ^* _1(x) /\psi ^*
_1 (x_1)$, so that
\be
\p \bar \psi _1 (q_2) = \p \psi ^* _1 (q_2) - (x_1 - q_1) \p \psi ^* _1 
(q_2) \p \psi ^* _1 (q_1) + \O (x_1 - q_1)^2 \, .
\ee
Combining all, (\ref{partq}) is given by $2  \p \psi ^* _1 (q_2)
f_{3/2} ^{(1)}(q_2) $ and we find (\ref{finamppq}) with 
$\X_i$ given by (\ref{peeque}).

\medskip

In deriving the above limit, we started from a formula
that had already combined the $\X_1$ and $\X_6$ contributions into their
sum $\X_1+\X_6$, which admits a finite limit. It is, however, important
to stress that neither $\X_1$ nor $\X_6$ by itself admits a limit as
$x_\alpha \to q_\alpha$. Indeed, the limit of $\X_1$ would correspond
to putting the supercurrent $S(x_\alpha)$ operator on top of the
superghost insertion $\delta (\beta (q_\alpha))$; but this limit does not
exist. Remarkably, the inclusion of the effects of the finite dimensional
gauge fixing determinants which result in $\X_6$ render the limit
well-defined. In particular, we obtain a well-defined interpretation of
the picture changing operator $Y(z)=\delta(\beta(z))S(z)$.
We view this intermediate result as one of the key successes of our
approach.

\end{document}